\documentclass[conference]{IEEEtran}

\usepackage{cite} 
\usepackage[T1]{fontenc}
\usepackage{algorithm}
\usepackage{algorithmic}

\ifx\pdfoutput\undefined
\usepackage{graphicx}
\else
\usepackage[pdftex]{graphicx}
\fi

\graphicspath{{./figs/},{./images/}{./plots/}}

\ifx\pdfoutput\undefined
\usepackage[hypertex]{hyperref}
\else
\usepackage[pdftex,hypertexnames=false]{hyperref}
\fi

\begin{document}

\title{On Irregular Interconnect Fabrics for Self-Assembled Nanoscale
Electronics}


\author{\authorblockN{Christof Teuscher}
\authorblockA{Los Alamos National Laboratory\\
CCS-1, MS-B287, 
Los Alamos, NM 87545, USA\\
E-mail: christof@teuscher.ch}}

\maketitle

\begin{abstract}
Nanoscale electronics and novel fabrication technologies bear unique
opportunities for self-assembling multi-billion component systems in a
largely random manner, which would likely lower fabrication costs
significantly compared to a definite ad hoc assembly. It has been
shown that communication networks with the small-world property have
major advantages in terms of transport characteristics and robustness
over regularly connected systems.  In this paper we pragmatically
investigate the properties of an irregular, abstract, yet physically
plausible small-world interconnect fabric that is inspired by modern
network-on-chip paradigms. We vary the framework's key parameters,
such as the connectivity, the number of switch blocks, the number of
virtual channels, the routing strategy, the distribution of long- and
short-range connections, and measure the network's transport
characteristics and robustness against failures. We further explore
the ability and efficiency to solve two simple toy problems, the
synchronization and the density classification task. The results
confirm that (1) computation in irregular assemblies is a promising
new computing paradigm for nanoscale electronics and (2) that
small-world interconnect fabrics have major advantages over local
CA-like topologies. Finally, the results will help to make important
design decisions for building self-assembled electronics in a largely
random manner.
\end{abstract}

\IEEEpeerreviewmaketitle

\section{Introduction and Motivation}
\label{sec:intro}
Despite important progress in recent years, nanoscale electronics is
still in its infancy and there is no consensus on what type of
computing architecture holds most promises.  Most of the effort in
nanotechnology has been focused in the past few years on developing
molecular devices that would eventually replace the traditional CMOS
transistor, but the development of higher-level computational
architectures for such devices always played a secondary role. As Chen
et al.  \cite{chen03} state, ``[i]n order to realize functional
nano-electronic circuits, researchers need to solve three problems:
invent a nanoscale device that switches an electric current on and
off; build a nanoscale circuit that controllably links very large
numbers of these devices with each other and with external systems in
order to perform memory and/or logic functions; and design an
architecture that allows the circuits to communicate with other
systems and operate independently on their lower-level details.''

The physical realization of computations from an abstract computing
machine is a challenging task, which is usually guided by a number of
major tradeoffs in the design space, such as the number and
characteristics of the resources available, the required performance,
the energy consumption, and the reliability. The lack of systematic
understanding of these issues and of clear design methodologies makes
the process still more of an art than of a scientific endeavor. The
appearance of novel and non-standard physical computing devices for
nanoscale and molecular electronics (such as for example array-based
\cite{dehon03} architectures or random assemblies of molecular gates
\cite{tour02}) only aggravates these difficulties.

In recent years, the importance of interconnects on chips has outrun
the importance of transistors as a dominant factor of chip performance
\cite{meindl03,ho01,davis01}. The ITRS roadmap \cite{itrs2003} lists a
number of critical challenges for interconnects and states that ``[i]t
is now widely conceded that technology alone cannot solve the on-chip
global interconnect problem with current design methodologies.'' The
major problems are related to delays of non-scalable global
interconnects and reliability in general, which leads to the
observation that simple scaling will no longer satisfy performance
requirements as feature sizes continue to shrink \cite{ho01}.

In this paper we pragmatically investigate a certain class of
irregular, physically plausible 3D interconnect fabrics, which are
likely to be easily and cheaply built by future self-assembling
processes for nanoscale electronics. We vary the framework's key
parameters, such as the connectivity, the number of switch blocks, the
number of virtual channels, the routing strategy, the distribution of
long- and short-range connections, and measure the network's transport
characteristic and robustness against failures. As a reference, we
will compare its performance with regular and nearest-neighbor
connected 2D and 3D cellular-automata-like fabrics. In addition to
previous work, we will also evaluate and compare the performance of
two toy tasks which are frequently used in the cellular automata
community, the synchronization and the density classification
task. The ability to solve a given task efficiently by means of a
certain interconnect topology has been a research topic since the
early age of parallel computing architectures.

The motivation for investigating alternative and more
biologically-inspired interconnects can be summarized by the following
observations: (1) long-range and global connections are costly and
limit system performance \cite{ho01}; (2) it is unclear whether a
precisely regular and homogeneous arrangement of components is needed
and possible on a multi-billion-component nanoscale assembly
\cite{tour02}; (3) ``[s]elf-assembly makes it relatively easy to form
a random array of wires with randomly attached switches''
\cite{zhirnov01}; and (4) building a perfect system is very hard and
expensive.

By using an abstract, yet physically plausible and
fabrication-friendly nanoscale computing framework, we will show that
interconnect fabrics with small-world-like \cite{watts98} properties
have major advantages in terms of performance and robustness over
purely regular and nearest-neighbor connected fabrics. We think that
the results will help to make important design decisions for building
self-assembled electronics in a largely random manner. Compared to
purely theoretical approaches, our framework provides more realistic
results.

The remainder of the paper is as following: Section \ref{sec:complex}
gives brief introduction to complex networks. The framework is
presented in Section \ref{sec:framework}. Sections
\ref{sec:experiments_a}, \ref{sec:experiments_b}, and
\ref{sec:experiments_c} describe various experiments and comparisons,
and Section \ref{sec:conclusion} concludes the paper.

\section{An Overview on Complex Networks}
\label{sec:complex}
Most real networks, such as brain networks \cite{sporns04,egueluz05},
electronic circuits \cite{cancho01}, the Internet, and social networks
share the so-called {\em small-world} (SW) property
\cite{watts98}. Compared to purely locally interconnected networks
(such as the cellular automata interconnect), small-world networks
have a very short average distance between any pair of nodes, which
makes them particularly interesting for efficient communication.

The classical Watts-Strogatz small-world network \cite{watts98} is
built from a regular lattice with only nearest neighbor
connections. Every link is then rewired with a {\em rewiring
  probability} $p$ to a randomly chosen node. Thus, by varying $p$,
one can obtain a fully regular ($p=0$) and a fully random ($p=1$)
network topology. The rewiring procedure establishes ``shortcuts'' in
the network, which significantly lower the average distance (i.e., the
number of edges to traverse) between any pair of nodes. In the
original model, the length distribution of the shortcuts is uniform
since a node is chosen randomly. If the rewiring of the connections is
done proportional to a power law, $l^{-\alpha}$, where $l$ is the wire
length, then we obtain a {\em small-world power-law network}. The
exponent $\alpha$ affects the network's transport characteristics
\cite{kozma05} and navigability \cite{kleinberg00}, which is better
than in the uniformly generated SW network. One can think of other
distance-proportional distributions for the rewiring, such as for
example a Gaussian distribution, which has been found between certain
layers of the rat's neocortical pyramidal neurons
\cite{hellwig00}. Studying the connection probabilities and the average
number of connections in biological systems, especially in neural
systems, can give us important insights on how nearly optimal systems
evolved in Nature under limited resources and various other physical
constraints.

In a real network, it is fair to assume that local connections have a
lower cost (in terms of resources required and delay) than
long-distance connections.  Physically realizing small-world networks
with uniformly distributed long-distance connections is thus not
realistic and distance, i.e., the wiring cost, needs to be taken into
account \cite{petermann05,petermann06,jespersen00}.

On the other hand, a network's topology also directly affects how
efficiently problems can be solved. For example, it has been shown
that both SW topologies \cite{tomassini05} as well as random
Erd\"os-R\'enyi topologies \cite{mesot05} have better performance than
regular lattices and are easier to evolve to solve the global
synchronization and density classification task. Thus, although rather
easy to realize, local-neighborhood networks are not generally
suitable for solving problems efficiently because of their poor global
transport characteristics. We will further address this in Section
\ref{sec:experiments_c}.

In this paper we are interested in networks with the small-world
property and a non-uniform distribution of the long-distance
connections because they present a realistic model of a
fabrication-friendly, self-assembled nano-scale interconnect fabric.

\section{Description of the Framework}
\label{sec:framework}
In order to compare representative regular nearest-neighbor and
irregular small-world interconnect fabrics, we use an abstract, yet
physically realistic system- and network-on-chip-like framework and an
evaluation methodology inspired by Pande et al. \cite{pande05}. We
will compare selected measures that are relevant for real systems.

The main challenge of interconnect fabrics---seen from a bird's eye
view---consists in transferring data between two points of the chip
with a minimal latency, minimal energy consumption, and maximal
reliability. This job can obviously be done in a wide variety of ways.
As opposed to the monolithic ad hoc interconnect networks used in
traditional chip design, we draw inspiration from recent {\em
  network-on-chip} (NoC) \cite{benini02,pande05} paradigms, which
transmit data in the form of packets on a routing network from a
source to the destination.

\begin{figure}
  \centering \includegraphics[width=0.35\textwidth]{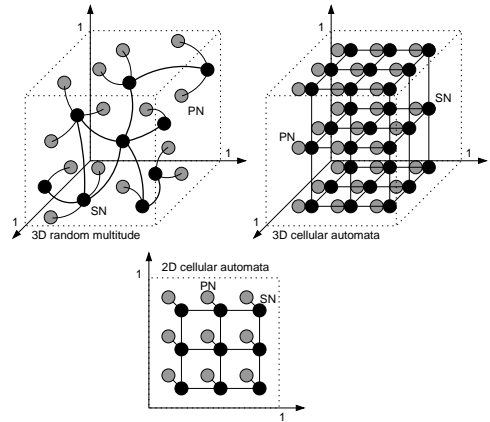}
  \caption{Top left: a random multitude (RM) example composed of
    processing nodes (PNs), switch nodes (SNs), and
    interconnections. Top right: a 3D CA-like architecture. Bottom: a
    2D CA-like architecture.}
  \label{fig:topo}
\end{figure}

\subsection{Regular 2D and 3D CA-like Architectures}
Both 2D and 3D {\em cellular automata} (CA) like architectures are
used as representatives of regular nearest-neighbor interconnect
fabrics. The basic system-on-chip-like architecture is composed of
programmable computing elements, called {\em processing nodes} (PNs),
and of an associated switch-based interconnect fabric, which is itself
composed of {\em switch nodes} (SNs) and bi-directional point-to-point
interconnects. Both PNs and SNs might be considered as simple IP
blocks. Each SN can execute and transmit in parallel messages on $C$
different virtual channels to its neighbors (see e.g. \cite{pande05}
for more details about the concept of virtual channels). We use an
unfolded version (see Figure \ref{fig:topo}, top right and bottom),
called {\em CLICH\'E} in \cite{pande05}, since folding requires
long-distance connections. The PNs are regularly arranged in the 2D or
3D Euclidean space inside a unitary square, respectively cube. The
number of PNs is equal to the number of SNs, and each PN is connected
to its associated SN by a single connection of $0.01$ unit length. For
our purposes, the PNs are able to send and receive messages, whereas
the SNs perform routing only.

\subsection{The Irregular Random Multitude (RM)}
In a {\em random multitude} (RM), both PNs and SNs are randomly
arranged in 3D space, as illustrated in Figure \ref{fig:topo} (top
left). To make comparisons with the CA-like architectures easier, we
assume that each PN is connected to the nearest SN in space by a
single connection only.  In this paper, we explore two different
distributions for establishing long-distance connections: (1)
power-law and (2) Gaussian. In case of a power-low distribution, the
SNs are connected among themselves by a small-world power-law network
\cite{petermann05,petermann06,jespersen00} with average connectivity
$SN_k$, i.e., each node establishes connections with its neighbors
proportional to $l^{-\alpha}$, where $l$ is the Euclidean distance
between the two SNs in question. Thus, the bigger $\alpha$, the more
local the connections. For $\alpha = 0$, we obtain the original
Watts-Strogatz SW topology. In case of a Gaussian distribution, the
connections with the neighbors are established proportional to
$f(l,\sigma) = \frac{1}{\sigma\sqrt{2\pi}}e^{-\frac{l^2}{2\sigma^2}}$,
where $l$ is the distance between the two SNs in question and $\sigma$
is the standard deviation. Thus, the smaller $\sigma$, the more local
the connections. For $\sigma = \infty$, we obtain the original
Watts-Strogatz SW topology. Compared to the power-law distribution,
the Gaussian distribution has a higher proportion of local
connections.

Algorithm 1 summarizes the construction of a random multitude from an
algorithmic point of view. $\delta$ is used to have some variability
in the connectivity around the mean value.

\begin{algorithm}[htb]
\begin{small}
  \label{alg:rmconstr}
  \caption{Construction of a 3D random multitude (RM)}
    \begin{algorithmic}[1]
      \STATE Randomly position $N$ processing nodes within a $1
        \times 1 \times 1$ unit cube (at distinct positions).
      \STATE Randomly position $S$ switch nodes within a $1
        \times 1 \times 1$ unit cube (at distinct positions).
      \FOR{each processing node $n$}
        \STATE Connect  $n$ to its nearest switch node $s$.
      \ENDFOR
      \FOR{each switch node $s$}
        \STATE Draw $\delta$ from a probability distribution with mean $0$ 
               and standard deviation $\sigma$, e,g, a Gaussian distribution. 
        \STATE Switch node (SN) connectivity $SN_k$.
        \FOR{$k = 1$ to $SN_k \pm \delta$}
          \STATE Connect $s$ to a neighboring switch node with probability 
            proportional to a connection probability function $f(l,\sigma)$, 
            where $l$ is the Euclidean distance between two nodes and 
            $\sigma$ the standard deviation.
        \ENDFOR
      \ENDFOR
    \end{algorithmic}
\end{small}
\end{algorithm}

\subsection{Physical Realization of a Random Multitude}
\label{sec:realization}
There exists an abundance of abstract computing models which are
either hard or impossible (e.g., when infinite resources or time is
involved) to physically realize. Despite progress, the fabrication of
ordered 3D hierarchical structures remains very challenging
\cite{zhang06}. Because of fewer physical constraints, we argue that
computing architectures that are ``assembled'' in a largely random
manner are easier and cheaper to build than highly regular
architectures, such as crossbars or CA-like assemblies, which usually
require a perfect or almost perfect establishment of the
connections. Self-assembly, for example, is particularly well suited
for building random structures \cite{zhirnov01}. Power-law
connection-length distributions have been observed in many systems
created through self-organization, such as the human cortex or the
Internet, and they can be considered ``physically realizable''
\cite{petermann06}. Such topologies evolve naturally in Nature because
of the cost associated with long distance connections. There is very
little work about computing architectures with irregular assemblies of
connections and components. Tour et al.  \cite{tour02}, for example,
explored the possibility of computing with randomly assembled, easily
realizable molecular switches, that were only locally interconnected
though. On the other hand, Hogg et al. \cite{hogg06} present an
approach to build reliable circuits by self-assembly with some random
variation in the connection location.

Designing nanoscale interconnects is guided by a number of dependent
major tradeoffs: (1) the number of long(er)-distance connections, (2)
the physical plausibility, and (3) the efficiency of
communication. Being able to physically realize a RM is crucial for
the success of such an unconventional architecture. Although we do not
provide any concrete solution, a plausible approach shall be sketched
here. We believe that a random multitude would be best realized in a
hybrid way today, where the PNs and SNs are for example made of
current (nanoscale) silicon. The interconnect fabric would then be
gradually self-assembled using nanoscale techniques such as directed
assembly \cite{ye06} by means of electrodeposition or vapor
deposition, or any other suitable technique. To obtain a power-law
distribution of connection lengths, one might imagine fabricating a
large amount of wires first, whose lengths follow a power law
distribution.  In a second step, they would be immersed in a solution
together with the nodes, randomly aligned (e.g., by means of electric
fields), and soldered as described in \cite{ye06}. Note that current
nanowires tend to be fairly short because of a high resistance and
probability of breaks, which will limit the number of long-distance
connections today.

\section{Experiments A: Exploring Parameters}
\label{sec:experiments_a}
The goal of this initial experiment is to vary certain key parameters
and thus to be able to make better design choices. We would like to
answer questions such as: (1) What is the right connectivity? (2) What
is the right distribution of local and global connections? (3) How
does the number of virtual channels affect throughput? (4) How does
the number of switch nodes (SNs) affect performance? As we will see,
most of these questions cannot be reduced to a single value because of
the various tradeoffs involved.

\subsection{Methods}
We have systematically explored the parameter space of $\alpha$,
$\sigma$, $S$, and $SN_k$ of the framework as described above, with
both a power-law and a Gaussian distribution of long-distance
connections. The number of processing nodes was fixed to $N=125$ for
all experiments. As a simplification, all buffer sizes are considered
unlimited. All nodes are updated asynchronously. Our simulations use a
simplistic random traffic model, which generates a message with a
random destination with a certain probability {\em trLoad} in each
PN. If $trLoad = 1$, a massage will be generated in each node at each
update. In the following experiments, we used a traffic load of
$trLoad = 0.1$. We have also implemented, random, shortest path, and
ant routing \cite{caro98}, but will focus on random routing in this
section since the results are more pronounced and illustrative. Also,
due to the similarity between the two distributions, we'll only
present results for the power-law distribution in the following
results section. The Gaussian distribution will be compared in Section
\ref{sec:experiments_b}.

\subsection{Results}
\begin{figure}
  \centering \includegraphics[width=0.43\textwidth]{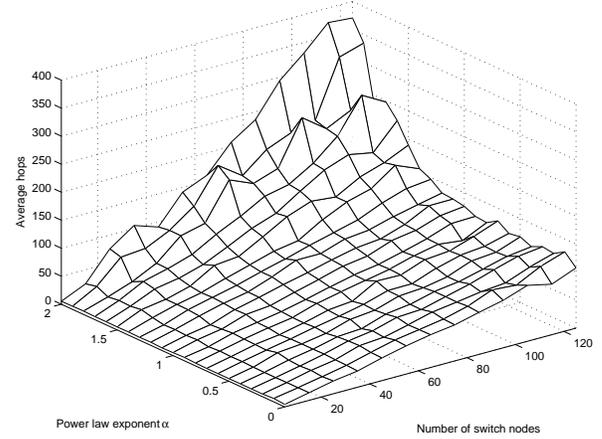}
  \caption{Average number of hops as a function of the power-law
    exponent $\alpha$ and the number of switch nodes $S$. The
    distribution of long-distance connections is uniform if
    $\alpha=0$. $N=125$.}
  \label{fig:surf_alpha_switch_hops}
\end{figure}

\begin{figure}
  \centering \includegraphics[width=0.43\textwidth]{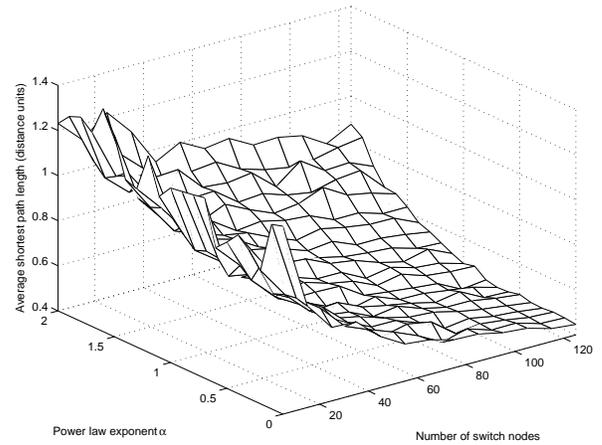}
  \caption{Average shortest path length as a function of the power-law
    exponent $\alpha$ and the number of switch nodes $S$. $N=125$.}
  \label{fig:surf_alpha_switch_path}
\end{figure}

Figure \ref{fig:surf_alpha_switch_hops} shows that the average number
of hops for a message to take from a any source to any destination
increases with an increasing number of switch nodes and a decreasing
number of global connections if $S$ gets bigger. On the other hand, as
Figure \ref{fig:surf_alpha_switch_path} shows, the smaller the number
of switch nodes, the longer the average shortest path length. Thus,
depending on what the network needs to be optimized for (i.e., lower
number of hops or shorter average path length), one can make the
appropriate choice for the number of switch nodes. Obviously, the
amount of hardware resources and the volume required will also come
into play in reality.

\begin{figure}
  \centering \includegraphics[width=0.43\textwidth]{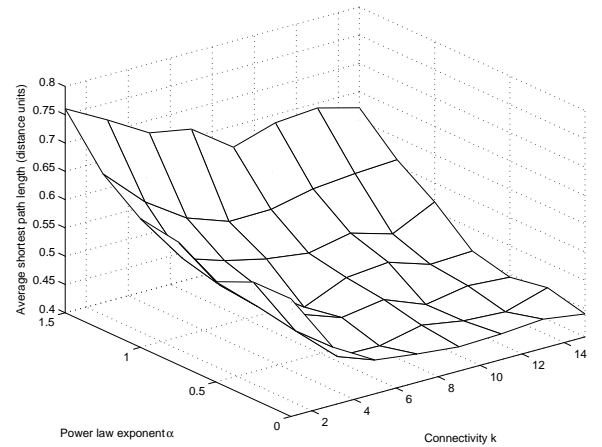}
  \caption{Average shortest path as a function of the power-law
    exponent $\alpha$ and the connectivity $SN_k$. $N=S=125$.}
  \label{fig:surf_alpha_k_path}
\end{figure}

Figure \ref{fig:surf_alpha_k_path} illustrates that the higher the
connectivity and the more global the connectivity (i.e., $\alpha =
0$), the lower the average shortest path length. Due to a lack of
space for more figures, the main results shall be summarized:

\begin{itemize}
  \item A higher switch node connectivity decreases both the average
    latency and the average number of hops. The throughput is only
    slightly improved.

  \item The higher the number of switch nodes $S$, the higher the
    number of hops and the higher the average latency. The lower $S$,
    the higher the average path length and the higher the throughput
    (measured in messages/update/switch node).

  \item The higher the number of virtual channels $C$, the higher the
    node throughput (within the limits of the capacity of the physical
    links) and the lower the average latency. The average shortest
    path length is not affected by $C$.

\end{itemize}

\subsection{Discussion}
There are no ``optimal'' values for connectivity, the number of switch
nodes, and the number of virtual channels. Instead, choosing the right
values is a matter of dependent tradeoffs in the design space. Local
connections are very interesting from an implementational point of
view, but offer diminished global transport characteristics only,
which directly affects the efficiency of problem solving. Adding a few
long(er)-distance connections proportional to the distance between the
nodes is physically plausible and greatly improves the overall
transport characteristics (i.e., small-world property) as well as the
robustness, as we will see in the next section.

In the following experiments, we used $6$ virtual channels and $N=S$
in order to be able to compare the results with the 3D CA-like
arrangement.

\section{Experiments B: Comparison with CA-like Interconnects}
\label{sec:experiments_b}
We performed a number of experiments to compare and contrast the
different interconnect architectures as described in Section
\ref{sec:framework}.

\subsection{Methods}
As a simple showcase, we assume that each SN can only perform either
random (RR) or shortest path (SPR) routing. Many other and more
efficient routing techniques exist, but we consider these two as
simple representatives of the least and the most effective methods. To
be able to compare the results with the CA-like topologies, we keep
the number of SNs and PNs equal in the random multitude
architecture. We measure the following performance metrics: (1)
average message latency (in clock cycles); (2) the average shortest
path length (in distance units); (3) the average number of hops; (4)
and the throughput (in messages/number of updates/switch node). For
all experiments in this section, we used $S=N=64$, $6$ virtual
channels per node (i.e., a 3D CA-node could send a message into all
directions simultaneously), an average SN connectivity of $SN_k = 6$,
an exact PN connectivity of $1$, and a traffic load of $trLoad =
0.1$. For our purposes, we kept all these parameters constant as a
detailed analysis would have been beyond the scope of this paper.

\subsection{Results}
\begin{figure}
  \centering \includegraphics[width=0.43\textwidth]{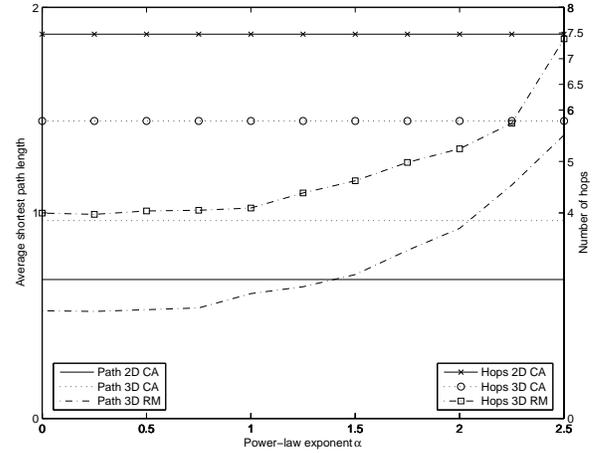}
  \caption{Average length of shortest path and number of hops as a
    function of $\alpha$. Shortest path routing (SPR), average values
    over 2 runs.}
  \label{fig:pathhops}
\end{figure}

Figure \ref{fig:pathhops} shows the average length of the shortest
paths (left y-axis) between each pair of PNs and the number of hops
(right y-axis) as a function of the power-law exponent $\alpha$. As
one can see, the average shortest path gets shorter the smaller
$\alpha$, i.e., the more long-distance connections exist. The effect
on the number of hops is identical.

According to \cite{petermann06}, a network is in a small-world regime
if $\alpha < 2D$, where $D$ is the dimension of the original
lattice. Compared to \cite{petermann06}, we allow multiple connections
and the nodes are randomly arranged, but the above equation should
still approximately held for $D=3$.  We have chosen $\alpha = 1.25$
for the following experiments since with this value, the RM performs
just better than a 3D CA.
 
\begin{figure}
  \centering \includegraphics[width=0.43\textwidth]{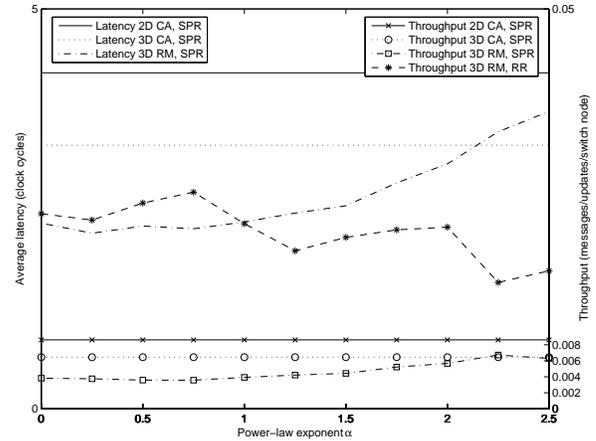}
  \caption{Latency and throughput as a function of $\alpha$.  Shortest
    path routing (SPR), random routing (RR), average values over $2$
    runs. $\alpha = 0$: uniform distribution of long-distance
    connections, $S=N=64$.}
  \label{fig:latthrough}
\end{figure}

Figure \ref{fig:latthrough} shows both latency and throughput as a
function of the power-law exponent $\alpha$. Due to the long-distance
connections, the random multitude has a lower latency than both 3D and
2D local-neighborhood interconnects. Rather surprisingly, the
throughput of the random multitude is the worst if one uses shortest
path routing (SPR). This is because SPR uses the same nodes for
numerous paths and thus creates more congestion because of the limited
number of channels per node. If one uses random routing (RR), as also
shown in Figure \ref{fig:latthrough}, the random multitude performs
best because the SNs are used more evenly and there is thus less
congestion. In reality, a routing algorithm which also considers
traffic and queue-lengths should be used.

\begin{figure}
  \centering \includegraphics[width=0.43\textwidth]{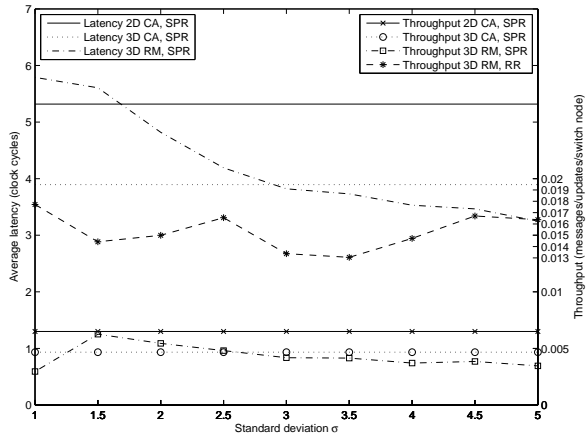}
  \caption{Latency and hops as a function of the standard deviation
    $\sigma$ of a Gaussian distribution. Shortest path (SPR) and
    random routing (RR). $\sigma = \infty$: uniform distribution of
    the connections. The break at $\alpha=0$ is due to a partly
    disconnected graph because of the local connectivity.}
  \label{fig:gau_latthrough}
\end{figure}

Figure \ref{fig:gau_latthrough} shows the same information as Figure
\ref{fig:pathhops}, but for a Gaussian distribution with standard
deviation $\sigma$. The bigger $\sigma$, the more uniform---and thus
global---the connectivity. $\sigma = \infty$ corresponds to the
original Watts-Strogatz model. In terms of absolute values, the
latency is worse in case of a Gaussian distribution, mainly because
the connectivity is more local. The throughput values are similar for
both random and shortest path routing.

\begin{figure}
  \centering \includegraphics[width=0.43\textwidth]{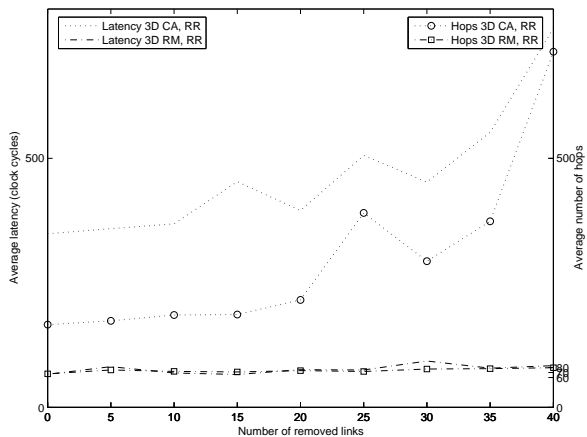}
  \caption{Latency and hops as a function of the number of removed
    links between SNs. Random routing (RR), $\alpha = 1.25$, average
    values over $4$ runs, $S=N=64$, power-law distribution.}
  \label{fig:dellatthrough}
\end{figure}

Finally, Figure \ref{fig:dellatthrough} illustrates what happens when
a certain number of links is removed randomly. The latency of the
random multitude is lowest and is basically unaffected by the random
removal of a rather small number of links. We used random routing (RR)
to illustrate an extreme case. The 2D CA is not shown because it
performs much worse than both the 3D CA and the RM. As one can see,
the number of hops is affected by the link removal in a similar way
than the latency. The results are similar for a Gaussian distribution.

\subsection{Discussion}
We have seen that small-world power-law networks perform better and
are more robust than 2D and 3D local-neighborhood interconnects in our
framework. The reason for this are the few longer-distance
connections, which provide short-cuts in the network. Compared to a
random network (which also has small-world properties), the
small-world power-law networks we use are more fabrication friendly
and very resource economical because they only use a very limited
number of longer-distance connections. As Jespersen and Blumen
\cite{jespersen00} state, networks with $\alpha < 2$ differ
significantly from those of regular lattices, which our experiments
confirm. We have also seen that small-world power-law networks are
very robust with respect to link deletions. Petermann and De Los Rios
\cite{petermann05} have further shown that the mean distance increases
by removing links and that the system becomes more fragile as $\alpha$
increases. Finally, as our experiments show, 3D local-neighborhood
interconnects require a lower number of hops and have a lower average
latency than their 2D counterpart. 3D fabrics thus presents an obvious
solution to interconnect problems. It has been shown elsewhere
\cite{davis01} that the average total wire lengths are shorter and
that fewer and shorter semi-global and global wires are required in 3D
interconnects.

\section{Experiments C: Solving Problems}
\label{sec:experiments_c}
In this last set of experiments, we are interested in evaluating the
performance of solving two ``toy problems'', which are well known in
the area of cellular automata (CA): the {\em synchronization} and the
{\em density classification} task. Both of these ``global'' tasks are
mostly trivial to solve if one has a global view on the entire system
(i.e., if one has access to the state of all nodes at the same time),
but are non-trivial to solve for locally connected cellular automata
or random boolean networks (RBNs). Although they are commonly called
toy problems, especially the synchronization task has actually many
real-world applications, such as for example in sensor networks, where
one cannot assume global synchronization and global signals, and thus
requires special mechanisms \cite{li06}.

In the {\em density classification task}, each node of a cellular
system must decide whether or not the initial configuration of the
automaton contains more than $50\%$ of $1$s. In this context, the term
``configuration'' refers to an assignment of the states $0$ or $1$ to
each cell of the system (i.e., there are $2^N$ possible initial
configurations). The desired behavior of the automaton is to have all
of its cells set to $1$ if the initial density of $1$s exceeded $1/2$,
and all $0$s otherwise.  The density classification task was studied
by many people, e.g.,
\cite{mitchell94,capcarrere01,tomassini02:cs,mesot05,sipper98:ijmpc},
in various forms, including non-uniform CAs, asynchronous CAs, and
non-standard architectures.

The {\em synchronization task} (also called {\em firefly task}) for
synchronous CAs was introduced by Das et al.  \cite{dasetal95} and
studied among others by Hordijk \cite{hordijk96} and Sipper
\cite{sipper97:book}. In this task, the two-state one-, two-, or
higher-dimensional automaton, given any initial configuration, must
reach a final configuration within $M$ time steps, that oscillates
between all $0$s and all $1$s on successive time steps. The whole
automaton is then globally synchronized.

Here, we use slightly modified versions of the two tasks that were
adapted to our framework.

\subsection{Methods}
For the synchronization task, we assume that each processing node (PN)
in our framework contains an oscillator which frequency is specified
by a number between $0 \leq f_{osc} \leq 1$. The modified task then
consists to find a common frequency for all oscillators. The algorithm
is inspired by the {\em averaging algorithm} as described in
\cite{li06}. Each processing node state is initialize to a random
value from the interval $[0$,$1]$ before it repeatedly performs the
following steps in an asynchronous manner: (1) send current oscillator
frequency to a random PN; (2) if the current node $i$ receives a
message from any other PN $r$, then average own oscillator $f_i$ with
neighbor frequency $f_r$; (3) set own oscillator to this frequency
$f_i = \frac{f_i + f_r}{2}$; and (4) also send it to a new random PN.

The density classification task is implemented in a similar way. Each
node can have a value $d$ from the interval $[0,1]$ and is initialized
randomly. If more then $50\%$ of the values are bigger then $0.5$, we
want all nodes to converge towards $1$, otherwise towards $0$. Each
node thus repeatedly performs the following steps in an asynchronous
manner after the initialization: (1) send current node value $d$ to a
random PN; (2) if the current node $i$ receives a message from any
other PN $r$, then average $d_i$ with $d_r$; (3) set $d_i$ to this
value, $d_i = \frac{d_i+d_r}{2}$; (4) if $d > 0.5$, then send $d_i +
\frac{1-d_i}{2}$ to a random PN, otherwise send $d_i - \frac{d_i}{2}$.

There are obviously numerous (also more efficient) ways to solve these
two tasks, but here we are interested in an illustrative comparison
rather than in the absolute performance values and limits. We thus
compared how these two simple algorithms perform on the investigated
interconnect fabrics using random routing.

\subsection{Results}
\begin{figure}
  \centering \includegraphics[width=0.43\textwidth]{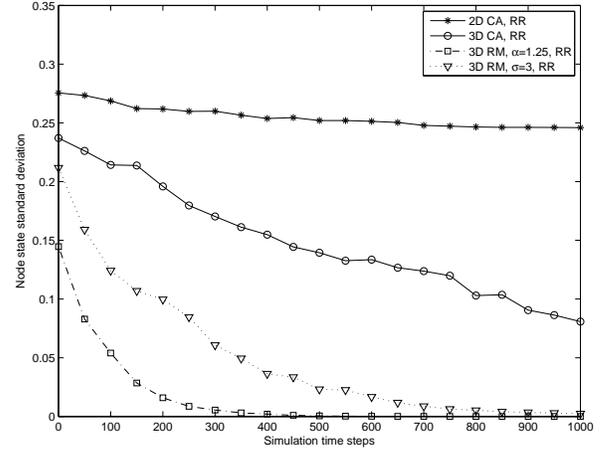}
  \caption{Performance of the synchronization task. The smaller the
    standard deviation of the node state values, the better the nodes
    are synchronized. The initial values depend on the randomly
    initialized network state.}
  \label{fig:sync_std}
\end{figure}

\begin{figure}
  \centering \includegraphics[width=0.43\textwidth]{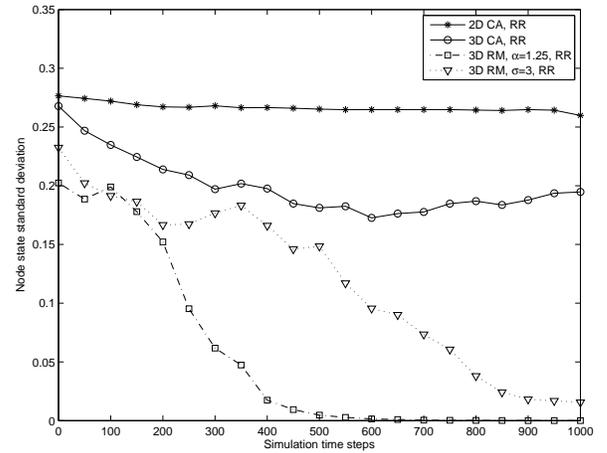}
  \caption{Performance of the density classification task. The smaller
    the standard deviation of the node state values, the better the
    density classification task is solved. The initial values depend
    on the randomly initialized network state.}
  \label{fig:dens_std}
\end{figure}

Figures \ref{fig:sync_std} and \ref{fig:dens_std} show the performance
of the synchronization and the density classification task
respectively. The smaller the standard deviation of the node state
values, the better the nodes are synchronized and the more successful
the density task is solved since all nodes have converged to the same
values. We used $N=S=125$ for both the random multitude and
the 3D CA and $N=S=121$ for the 2D CA.

As one can see, the small-world random multitude with a power-law
distribution of the connections performs best (e.g, convergence
towards a common value for all nodes is fastest) for both tasks,
before the Gaussian distribution, the 3D CA, and the 2D CA. The
results are similar, but convergence is much faster if one uses
shortest path routing instead of random routing.

\subsection{Discussion}
It has been shown elsewhere that irregular small-world interconnects
perform better on both the synchronization (e.g.,
\cite{guclu06,mesot05,hong04} and many others) and the density
classification task (e.g., \cite{mesot05}) than purely locally
interconnected topologies. However, the frameworks and assumptions
used in each approach are somehow different and sometimes not
straightforward to compare. The results of our toy framework merely
confirm what has been found theoretically elsewhere and in our two
previous experiments, namely that the excellent transport
characteristics (i.e., short characteristic path length, small
latency, etc.) also helps to efficiently solve tasks, especially tasks
which require a lot of global communication. From an evolutionary
perspective, this is also the reason why most natural networks,
e.g. the brain \cite{hellwig00,egueluz05,sporns04}, have evolved with
the small-world and scale-free property.

\section{Conclusion}
\label{sec:conclusion}
We have investigated in a pragmatic way several relevant metrics of
both regular and irregular, realistic system-on-chip-like computing
architectures for self-assembled nanoscale electronics, namely 2D and
3D local-neighborhood as well as two random small-world interconnects
with different distributions for long-distance connections. The
small-world architectures are both physically plausible, could likely
be built very economically by self-assembling mechanisms, possess
great transport characteristics, and are robust against link
failures. While regular and local-neighborhood interconnects are
easier and more economical to build than interconnects with lots of
global or semi-global long-distance connections, we have seen in the
last Section that they are not as efficient for global communication,
which is very important and directly affects how efficient problems
can be solved in general. Small-world networks with a uniform
distribution of long-distance connections or pure random networks, on
the other hand, are not physically plausible because one has to assume
an increasing cost for connections with distance.  As our results have
shown by means of our simplistic, yet realistic framework, small-world
power-law interconnects offer a unique balance between performance,
robustness, physical plausibility, and fabrication friendliness. In
addition, it has been shown that adaptive routing---which we haven't
explored here---is very efficient on small-world power-law graphs
\cite{kleinberg00}.

We believe that computation in random self-assemblies of components
(e.g., \cite{tour02}) is a highly appealing paradigm, both from the
perspective of fabrication as well as performance and robustness. This
is obviously a radical new technological and conceptual approach with
many open questions. For example, there are basically no methodologies
and tools that would allow (1) to map an arbitrary architecture on a
randomly assembled physical substrate, (2) to do arbitrary
computations with such an assembly, and (3) to systematically analyze
performance and robustness within a rigorous mathematical
framework. There are also many open questions regarding the
self-assembling fabrication techniques.

Future work will concentrate on the computational aspects of such
assemblies and not solely on the interconnects, as in the present
work. We also plan to evaluate further measures, such as energy
consumption and area used, and to develop localized and adaptive
routing strategies.


\begin{thebibliography}{10}

\bibitem{itrs2003}
International technology roadmap for semiconductors ({ITRS}).
\newblock Semiconductor Industry Association, {\tt http://public.itrs.net},
  2003.

\bibitem{benini02}
L.~Benini and G.~de~Micheli.
\newblock Networks on chips: A new {SoC} paradigm.
\newblock {\em IEEE Computer}, 35(1):70--78, 2002.

\bibitem{capcarrere01}
M.~Capcarrere and M.~Sipper.
\newblock Necessary conditions for density classification by cellular automata.
\newblock {\em Physical Review E}, 6403(3):6113--6117, 2001.

\bibitem{caro98}
G.~Di Caro and M.~Dorigo.
\newblock {AntNet: Distributed Stigmergetic Control for Communications
  Networks}.
\newblock {\em Journal of Artificial Intelligence Research}, 9:317--365, 1998.

\bibitem{chen03}
Y.~Chen, G.-Y. Jung, D.~A.~A. Ohlberg, X.~Li, D.~R. Steward, J.~O. Jepperson,
  K.~A. Nielsen, J.~F. Stoddart, and R.~S. Williams.
\newblock Nanoscale molecular-switch crossbar circuits.
\newblock {\em Nanotechnology}, 14:462--468, 2003.

\bibitem{dasetal95}
R.~Das, J.~P. Crutchfield, M.~Mitchell, and J.~E. Hanson.
\newblock Evolving globally synchronized cellular automata.
\newblock In L.~J. Eshelman, editor, {\em Proceedings of the Sixth
  International Conference on Genetic Algorithms}, pages 336--343, San
  Francisco, CA, 1995. Morgan Kaufmann.

\bibitem{davis01}
J.~A. Davis, R.~Venkatesan, A.~Kaloyeros, M.~Beylansky, S.~J. Souri,
  K.~Banerjee, K.~C. Saraswat, A.~Rahman, R.~Reif, and J.~D. Meindl.
\newblock Interconnect limits on gigascale integration ({GSI}) in the $21^{st}$
  century.
\newblock {\em Proceedings of the IEEE}, 89(3):305--324, 2001.

\bibitem{dehon03}
A.~DeHon.
\newblock Array-based architecture for {FET}-based nanoscale electronics.
\newblock {\em IEEE Transactions on Nanotechnology}, 2(1):23--32, 2003.

\bibitem{egueluz05}
V.~M. Egu\'eluz, D.~R. Chialvo, G.~A. Cecchi, M.~Baliki, and A.~V. Apkarian.
\newblock Scale-free brain functional networks.
\newblock {\em Physical Review Letters}, 94:018102, 2005.

\bibitem{guclu06}
H.~Guclu, G.~Korniss, M.~A. Novotny, Z.~Toroczkai, and Z.~R\'acz.
\newblock Synchronization landscapes in small-world-connected computer
  networks.
\newblock {\tt arXiv:cond-mat/0601058}, 2006.

\bibitem{hellwig00}
B.~Hellwig.
\newblock A quantitative analysis of the local connectivity between pyramidal
  neurons in layers 2/3 of the rat visual cortex.
\newblock {\em Biological Cybernetics}, 82:111--121, 2000.

\bibitem{ho01}
R.~Ho, K.~W. Mai, and M.~A. Horowitz.
\newblock The future of wires.
\newblock {\em Proceedings of the IEEE}, 89(4):490--504, 2001.

\bibitem{hogg06}
T.~Hogg, Y.~chen, and P.~J. Kuekes.
\newblock Assembling nanoscale circuits with randomized connections.
\newblock {\em IEEE Transactions on Nanotechnology}, 5(2):110--122, 2006.

\bibitem{hong04}
H.~Hong, B.~J. Kim, M.~Y. Choi, and H.~Park.
\newblock Factors that predict better synchronizability on complex networks.
\newblock {\tt arXiv:cond-mat/0403745}, 2004.

\bibitem{hordijk96}
W.~Hordijk.
\newblock The structure of the synchonizing-{CA} landscape.
\newblock Technical Report 96-10-078, Santa Fe Institute, Santa Fe, NM (USA),
  1996.

\bibitem{cancho01}
R.~Ferrer i~Cancho, C.~Janssen, and R.~V. Sole.
\newblock Topology of technology graphs: Small world patterns in electronic
  circuits.
\newblock {\em Physical Review E}, 64:046119, 2001.

\bibitem{jespersen00}
S.~Jespersen and A.~Blumen.
\newblock Small-world networks: Links with long-tailed distributions.
\newblock {\em Physical Review E}, 62(5):6270--6274, 2000.

\bibitem{kleinberg00}
J.~K. Kleinberg.
\newblock Navigation in a small world.
\newblock {\em Nature}, 406:845, 2000.

\bibitem{kozma05}
B.~Kozma, M.~B. Hastings, and G.~Korniss.
\newblock Diffusion processes on power-law small-world networks.
\newblock {\em Physical Review Letters}, 95:018701, 2005.

\bibitem{li06}
Q.~Li and D.~Rus.
\newblock Global clock synchronization in sensor networks.
\newblock {\em IEEE Transactions on Computers}, 55(2):214--226, 2006.

\bibitem{meindl03}
J.~D. Meindl.
\newblock Interconnect opportunities for gigascale integration.
\newblock {\em IEEE Micro}, 23(3):28--35, 2003.

\bibitem{mesot05}
B.~Mesot and C.~Teuscher.
\newblock Deducing local rules for solving global tasks with random {Boolean}
  networks.
\newblock {\em Physica D}, 211(1--2):88--106, 2005.

\bibitem{mitchell94}
M.~Mitchell, J.~P. Crutchfield, and P.~T. Hraber.
\newblock Evolving cellular automata to peform computations: Mechanisms and
  impediments.
\newblock {\em Physica D}, 75:361--391, 1994.

\bibitem{pande05}
P.~P. Pande, C.~Grecu, M.~Jones, A.~Ivanov, and R.~Saleh.
\newblock Performance evaluation and design trade-offs for network-on-chip
  interconnect architectures.
\newblock {\em IEEE Transactions on Computers}, 54(8):1025--1040, 2005.

\bibitem{petermann05}
T.~Petermann and P.~De~Los Rios.
\newblock Spatial small-world networks: A wiring-cost perspective.
\newblock {\tt arXiv:cond-mat/0501420}, 2005.

\bibitem{petermann06}
T.~Petermann and P.~De~Los Rios.
\newblock Physical realizability of small-world networks.
\newblock {\em Physical Review E}, 73:026114, 2006.

\bibitem{sipper97:book}
M.~Sipper.
\newblock {\em Evolution of Parallel Cellular Machines: The Cellular
  Programming Approach}.
\newblock Springer-Verlag, Heidelberg, 1997.

\bibitem{sipper98:ijmpc}
M.~Sipper, M.~S. Capcarrere, and E.~Ronald.
\newblock A simple cellular automaton that solves the density and ordering
  problems.
\newblock {\em International Journal of Modern Physics C}, 9(7):899--902,
  October 1998.

\bibitem{sporns04}
O.~Sporns, D.~R. Chialvo, M.~Kaiser, and C.~C. Hilgtag.
\newblock Organization, development, and function of complex brain networks.
\newblock {\em Trends in Cognitive Sciences}, 8(9):418--425, 2004.

\bibitem{tomassini05}
M.~Tomassini, M.~Giacobini, and C.~Darabos.
\newblock Evolution and dynamics of small-world cellular automata.
\newblock {\em Complex Systems}, 15(4):261--284, 2005.

\bibitem{tomassini02:cs}
M.~Tomassini and M.~Venzi.
\newblock Evolving robust asynchronous cellular automata for the density task.
\newblock {\em Complex Systems}, 13(3):185--204, 2002.

\bibitem{tour02}
J.~Tour, W.~L.~Van Zandt, C.~P. Husband, S.~M. Husband, L.~S. Wilson, P.~D.
  Franzon, and D.~P. Nackashi.
\newblock Nanocell logic gates for molecular computing.
\newblock {\em IEEE Transactions on Nanotechnology}, 1(2):100--109, 2002.

\bibitem{watts98}
D.~J. Watts and S.~H. Strogatz.
\newblock Collective dynamics of `small-world' networks.
\newblock {\em Nature}, 393:440--442, 1998.

\bibitem{ye06}
H.~Ye, Z.~Gu, T.~Yu, and D.~H. Gracias.
\newblock Integrating nanowires with substrates using directed assembly and
  nanoscale soldering.
\newblock {\em IEEE Transactions on Nanotechnology}, 5(1):62--66, 2006.

\bibitem{zhang06}
F.~Zhang and H.~Y. Low.
\newblock Ordered three-dimensional hierarchical nanostructures by nanoimprint
  litography.
\newblock {\em Nanotechnology}, 17:1884--1890, 2006.

\bibitem{zhirnov01}
V.~V. Zhirnov and D.~J.~C. Herr.
\newblock New frontiers: Self-assembly in nanoelectronics.
\newblock {\em IEEE Computer}, pages 34--43, January 2001.

\end{thebibliography}
\end{document}